# Realization of ultra-broadband IR up-conversion imaging

Xiaohong Li[1,6], Peng Bai[2,6], Siheng Huang[1], Xueqi Bai[1], Wenjun Song[1],

Xinran Lian[1], Cheng Hu[1], Zhiwen Shi[1], Wenzhong Shen[1], Yueheng Zhang[1,*],

Zhanglong Fu[3], Dixiang Shao[3], Zhiyong Tan[3], Juncheng Cao[3],

Cheng Tan[4], Gangyi Xu[4,5]

1. *Key Laboratory of Artificial Structures and Quantum Control, School of Physics and Astronomy, Shanghai Jiao Tong University, Shanghai 200240, China*

2. *Institute of Applied Physics and Computational Mathematics, Beijing 100088, China*

3. *Key Laboratory of Terahertz Solid-State Technology, Shanghai Institute of Microsystem and Information Technology, Chinese Academy of Sciences, Shanghai 200050, China*

4. *Key Laboratory of Infrared Imaging Materials and Detectors, Shanghai Institute of Technical Physics, Chinese Academy of Sciences, Shanghai 200083, China*

5. *Hangzhou Institute for Advanced Study, University of Chinese Academy of Sciences, Hangzhou 310024, China*

6. *These authors contributed equally*

*Corresponding authors, E-mail addresses: yuehzhang@sjtu.edu.cn

# Abstract

Ultra-broadband imaging devices with high performance are in great demand for a variety of technological applications, including imaging, remote sensing, and communications. An ultra-broadband up-converter is realized based on a p-GaAs homojunction interfacial workfunction internal photoemission (HIWIP) detector-light emitting diode (LED) device. The device demonstrates an ultra-broad response ranging from visible to terahertz (THz) with good reproducibility. The peak responsivity in the mid-infrared (MIR) region is 140 mA/W at 10.5 μm. The HIWIP-LED shows enormous potential for ultra-broadband up-conversion covering all infrared atmospheric windows, as well as the THz region, and the pixelless imaging of the MIR spot from $CO_2$ laser is further demonstrated. In addition，the proposed up-converter also performs as a near-infrared and visible detector under zero bias by using a bi-functional LED. Thanks to its ultra-wide response, the HIWIP-LED up-converter has great promise for stable, high-performance ultra-broadband pixelless imaging and multi-functional analysis systems.

## Introduction

Developing efficient ultra-broadband imaging systems, particularly for the mid-infrared (MIR)and terahertz (THz) ranges, where the fundamental molecular vibration fingerprint region is located, is of great interest due to the huge potential in a variety of applications such as astronomy, environmental monitoring, health, national security, and chemical diagnostics[1, 2]. As the terminal of the imaging system, the ultra-broadband photodetector has attracted great interest all around the world in the past few years.

Graphene has shown a broad absorption spectrum spanning from the ultraviolet to the mid-infrared[3, 4, 5, 6]. It is theoretically capable of detecting low photon energy due to its gapless band structure and is suitable for thermal-related detectors because of its unique properties of low electron heat capacity and weak electron-phonon interaction. Band-structure-engineering processed pure graphene (graphene quantum dot, graphene nanoribbon) based MIR photodetectors have achieved responsivity of ~0.4 A/W in MIR ranges (~10 μm) at low temperature[7]. Because of its appropriate energy band structure, black phosphorus (BP) is also a suitable 2D material for ultra-broadband detection[8, 9, 10]. At room temperature, photodetectors based on black arsenic phosphorus demonstrate a response range up to 8.2 μm, entering the second MIR atmospheric transmission window[9]. Although the performance of 2D material-based photodetectors (e.g., graphene, BP) was greatly improved in recent years, limitations in mass manufacturing and integration with existing readout circuits remain the two primary barriers to their widespread usage in ultra-broadband imaging systems. Broadband detection with organic photodetectors is also a prominent topic of research currently. The majority of organic compounds, however, are rather unstable. As a result, organic photodetectors are prone to degradation, particularly in harsh working conditions[11, 12, 13].

Semiconductor-based IR and THz photodetectors, on the other hand, are more advanced in the fabrication process and have proved their applicability in imaging systems. As a representative, quantum well infrared photodetectors (QWIPs) show the potential in MIR and FIR detection because of the high sensitivity, fast response speed, and high damage threshold. Imaging systems based on the pixel-based arrays (e.g., linear array and focal plane array (FPA)) are widely used today. Although QWIP-FPA has shown great success in MIR imaging, this strategy is difficult to extend to THz imaging[14]. Firstly, designing an optical coupling structure for imaging applications in the THz region is difficult due to the considerably longer wavelength detection[15, 16]. Furthermore, although QWIP can realize broadband detection by integrating multiple structures on the device, it is still difficult to design matching optical coupling structures. Secondly, THz QWPs must be operated at low temperatures (about 20 K) due to their low activation energy (around 10 meV)[17]. Because of the thermal mismatch between GaAs and Si, repetitive heating and cooling of a hybrid GaAs-based FPA-readout integrated circuit (ROIC) imaging device will damage the connection. As the temperature drops, the number of dead pixels increases[18]. All these problems make QWIP-FPA difficult to be a suitable choice for ultra-broadband detection and imaging.

To solve the problems, the notion of pixelless imaging based on semiconductor up-conversion is presented. The main idea is to design an up-conversion device that can convert low-energy photons to high-energy photons, allowing THz imaging to be transformed into well-studied near-infrared imaging. The up-conversion device in this imaging strategy is an entire large-size imaging cell. The detecting part of the up-conversion device receives and transmits the original picture, then the light-emitting diode (LED) part restores the image and emits near-infrared light, which is eventually 'seen' by a Si charge coupled device (CCD). As a result, there is no requirement for a connection between the device and the readout circuit, which effectively eliminates the thermal mismatch problem. It has been reported that a QWP-LED-based up-conversion device has successfully realized THz imaging, demonstrating the enormous potential of applying photon-type-detector-based up-conversion devices to THz imaging[18, 19].

However, the polarization selection rule in QWIP requires the detector to adopt a 45° edge coupled geometry to realize the optical coupling, which may naturally induce image distortion and stretching. Although using grating geometry to achieve normal incidence is a feasible alternative, the wavelength dependence of diffraction implies that grating detectors are not suitable for broadband detection. To avoid these problems, GaAs-based homojunction interfacial workfunction internal photoemission (HIWIP) detectors have recently been introduced into the THz imaging system[20, 21]. Because of the free-carrier absorption mechanism in the HIWIP-LED up-converter, the normal incidence is possible. In comparison to the QWP-LED, the HIWIP-LED has a broadband response (4.2 THz-20 THz) in the THz region and a low optimized NEP of 29.1 pW/Hz$^{1/2}$, indicating that it might be used in a wider range of applications. The ability of photodetectors to achieve broadband response is a distinctive feature brought by free-carrier absorption. However, this capability has been underestimated in previous studies, which focused only on a certain interval[20, 22, 23, 24]. In fact, the response bandwidth of detectors based on free-carrier absorption is wide enough to cover several IR atmospheric windows. This capability of realizing ultra-broadband response is the basis for multi-functional detection, which is urgently needed in many important areas.

In this work, we demonstrated the potential application of the HIWIP-LED in ultra-broadband detection and imaging. The p-GaAs HIWIP-LED up-converter shows an ultra-broadband response from the visible to THz (150 cm$^{-1}$) region, which is much broader than the bandwidth of 2D material-based photodetectors. The HIWIP detector part was well measured and shows high performance in the IR region. We also demonstrated a strong up-conversion signal of the HIWIP-LED up-converter in the MIR region (@10.6 μm) and successfully realized the MIR imaging of a $CO_2$ laser spot. Finally, it is found very interesting that the HIWIP-LED up-converter can also respond to the near-infrared and visible light at room temperature under zero bias, which originates from the interband transition of GaAs in the LED part.

## Results

**Device structure & up-conversion principle**

The schematic of the up-conversion device is shown in Fig.1a. The up-conversion device consists of a p-GaAs HIWIP detector and an LED designed for low-temperature operation. The entire device is grown on a semi-insulating GaAs substrate by molecular beam epitaxy (the detail of the wafer information can be found in the Method section).

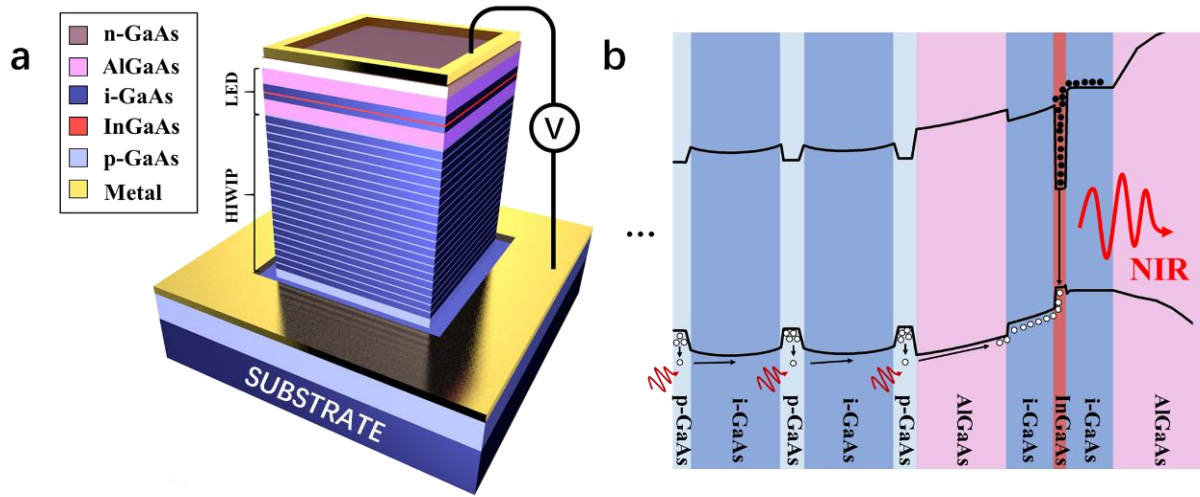

**Fig.1** Device structure and the schematic of the energy band diagram of the up-conversion device. **a** Structure of the up-conversion device, composed of a p-GaAs HIWIP detector and an LED operating at low temperature. The HIWIP detector is composed of 20 periods of a 150 Å p-GaAs and an 800 Å undoped GaAs barrier. The LED part consists of a thin layer of $In_{0.1}Ga_{0.9}As$ quantum well sandwiched between two AlGaAs/GaAs heterojunctions. **b** The schematic of the energy band diagram of the up-conversion device under reverse bias (with the top contact being grounded). The photo-excited carriers are generated in highly doped emitters because of free-carrier absorption (FCA) and inter-valence-band absorption (IVBA). Under reverse bias, the carriers are injected into the active region in the LED, recombine and emit NIR photons.

The schematic of the energy band diagram under reverse bias (with the top contact being grounded) is shown in Fig.1b. The photo-generated carriers result from free-carrier absorption (FCA) and inter-valence-band absorption (IVBA). On one hand, the free carriers in the highly doped emitter layers absorb the incident photon energy and overcome the interfacial work function caused by the band narrowing effect[21]. On the other hand, some carriers transit from the light/heavy hole band to the spin-orbit split-off band or transfer from the heavy hole band to the light hole band, thus contributing to the photocurrent[25]. The photon-excited carriers generated in the HIWIP detector part are injected into the

In$_{0.1}$Ga$_{0.9}$As quantum well of the LED part under reverse bias, recombine, and emit NIR photons that are detected easily by the Si CCD.

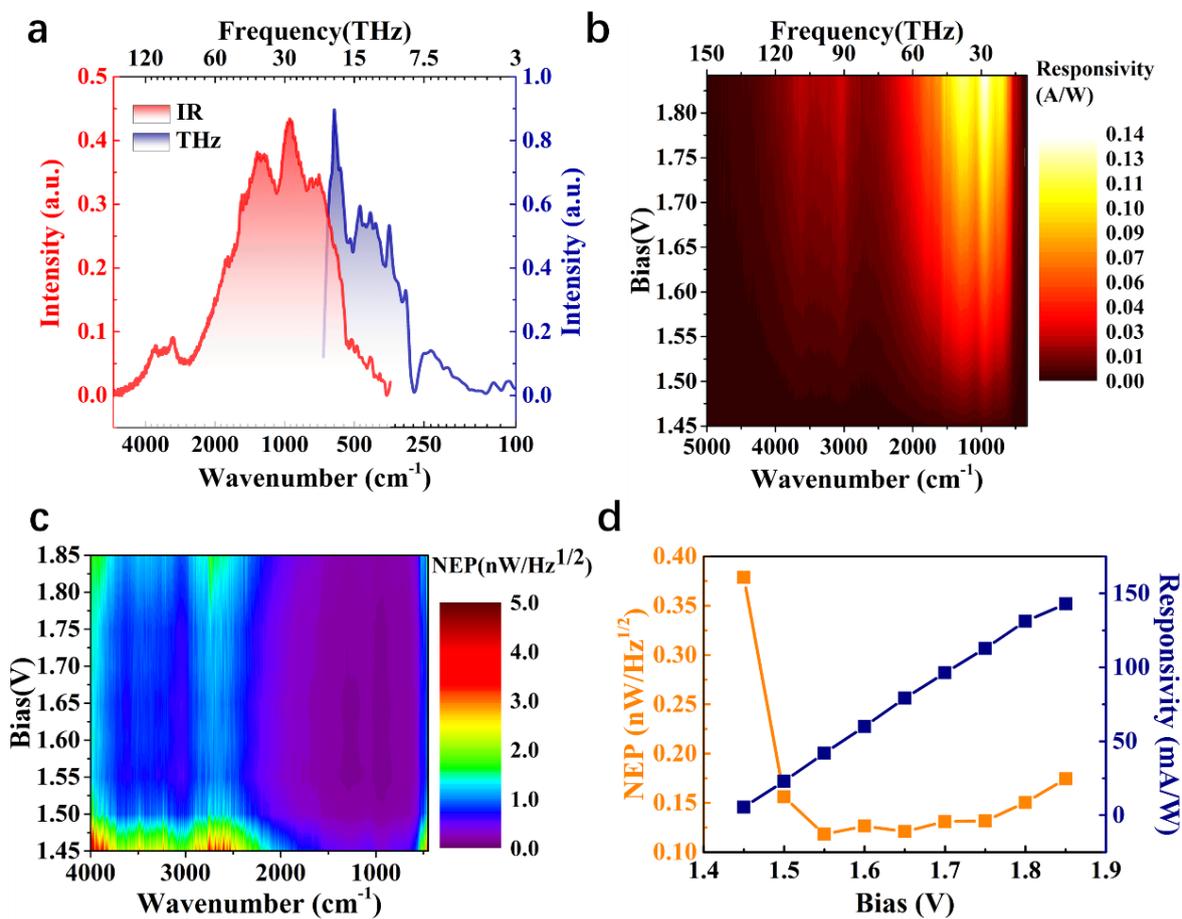

**Fig.2** Photoresponse of the HIWIP detector. **a** Photocurrent spectrum of the HIWIP-LED sample on an ultrabroad range from 150 cm$^{-1}$ to 5000 cm$^{-1}$ at 4.2 K in a high vacuum environment. The bias is set as 1.60 V, which is higher than the turn-on voltage (1.4 V) of LED. The red line shows the photocurrent spectrum of the HIWIP-LED sample in IR region. The blue line shows the photocurrent spectrum of the HIWIP-LED sample in the THz region. **b** Responsivity of the HIWIP-LED sample in MIR range at 4.2 K under different biases. The responsivity achieves a maximum of 0.14 A/W at ~10.5 μm when the bias is 1.842 V. **c** The NEP mapping of the HIWIP-LED sample in IR region at 4.2 K under different biases. The NEP reaches a minimum around 952 cm$^{-1}$, where the peak responsivity locates. **d** The minimum NEP and peak responsivity of the HIWIP-LED sample (~ 952 cm$^{-1}$) at 4.2 K under different biases.

**Performance of the HIWIP-LED in IR region**

The photocurrent spectrum of the HIWIP-LED sample on an ultra-broad range from 150 cm$^{-1}$ to 5000 cm$^{-1}$, covering the THz and MIR range, is shown in Fig.2a. The detailed information on the measurement can be found in the Method section. The excellent performance of p-GaAs HIWIP-LED for terahertz wave detection and terahertz up-conversion imaging (the blue line) has been demonstrated in detail in another research[20]. However, in practice, the FCA-based detectors have the ability to respond to broadband covering several IR atmospheric windows, the great potential of FCA-based detectors for applications in ultra-broadband detection has been grossly neglected before. In this paper, we will focus on the performance of the HIWIP-LED up-conversion device in the IR region and its possible applications. In the IR region (the red line), the p-GaAs HIWIP-LED shows a broad photoresponse from 350 cm$^{-1}$ to 5000 cm$^{-1}$ (2-28 μm). The peak position is at ~ 952 cm$^{-1}$ (10.5 μm). The broad-range response in the MIR region is due to the free-carriers absorption (FCA), while the peaks at 709 cm$^{-1}$ (14.1 μm), 952 cm$^{-1}$ (10.5 μm), 1316 cm$^{-1}$ (7.6 μm), and 3030 cm$^{-1}$ (3.3 μm) are produced by inter-valence-band-absorption (IVBA) including the hole transitions from the light/heavy-hole band to the spin-orbit split-off band and from heavy-hole band to light-hole band[25].

The responsivity of the HIWIP-LED sample under different biases at 4.2 K is shown in Fig.2b. The biases are chosen to be higher than the turn-on voltage (1.4 V) of LED. The responsivity shows an evident dependence on the bias voltage, which increases significantly with increasing bias. The responsivity achieves a maximum of 0.14 A/W at ~ 952 cm$^{-1}$ (10.5 μm) when the bias is 1.842 V.

The noise of the sample is evaluated using the dark current measured at 4.2 K. The noise equivalent power (NEP) is calculated using the evaluated noise and the responsivity obtained under different biases. We evaluate the overall NEP of the united imaging system consisting of the HIWIP-LED up-converter and the Si CCD. According to a developed noise theory of the up-conversion detectors[26], the noise of the united system is composed of three parts: the noise of the HIWIP detector, the noise of the LED, and the noise of the Si CCD, which is expressed as:

$$\begin{aligned} i_n^2 = & \ (\eta_{S_i}\eta_{LED})^2(4eg_{HW}i_{bg}\Delta f + 4eg_{HW}i_{dark}\Delta f) \\ & +(\eta_{Si}\eta_{LED})^2(2ei_{bg}\Delta f + 2ei_{dark}\Delta f) \\ & +2e\Delta f \eta_{Si}\eta_{LED}(i_{bg} + i_{dark}) + 2ei_{dark,Si}\Delta f \end{aligned}$$

where $\eta_{S_i}$ is the quantum efficiency of the Si CCD at the peak luminescence wavelength of the LED part; $\eta_{LED}$ is the external quantum efficiency of the LED part, which is measured to be ~2.4%, mainly restricted by the light extraction efficiency[20]; e is the elementary charge; $g_{HW}$ is the gain coefficient of the HIWIP part; $i_{bg}$ is the photocurrent caused by the 300 K background radiation; $i_{dark}$ is the dark current of the device; $\Delta f$ is the system measurement bandwidth; $i_{dark,Si}$ is the dark current of the Si CCD. The responsivity of the united imaging system is $R_{imag} = R\eta_{LED}\eta_{Si}hc/\lambda_{out}e$. Therefore, we can easily find that the NEP of the united up-conversion imaging system is:

$$\begin{aligned} NEP = \frac{i_n}{R_{imag}} = \frac{e\lambda_{out}}{hcR} & \Big[(4eg_{HW}i_{bg}\Delta f + 4eg_{HW}i_{dark}\Delta f) \\ & +(2ei_{bg}\Delta f + 2ei_{dark}\Delta f) \\ & + \frac{2e\Delta f}{\eta_{LED}\eta_{Si}}(i_{bg} + i_{dark}) + \frac{2ei_{dark,Si}\Delta f}{(\eta_{LED}\eta_{Si})^2}\Big]^{1/2} \end{aligned}$$

The generation-recombination noise is considered to be dominant in the up-conversion process. While the dark current of the commercial Si CCD can be suppressed to an extremely low level, the last term can also be neglected. The evaluated NEP is shown in Fig.2c. The NEP reaches a minimum around 952 cm$^{-1}$ (10.5 μm), where the peak responsivity locates. The minimum NEP and peak responsivity of the HIWIP-LED sample at 4.2 K under different biases is shown in Fig.2d. Values smaller than 150 pW/Hz$^{1/2}$ can be reached.

**Optical up-conversion and pixelless imaging**

The LED part in the HIWIP-LED sample is specially designed to operate at low temperatures. It consists of two AlGaAs/GaAs heterojunctions and a 9 nm $In_{0.1}Ga_{0.9}As$ quantum well sandwiched between them. The AlGaAs/GaAs heterojunction between the $In_{0.1}Ga_{0.9}As$ quantum well and the HIWIP part is designed to be intrinsic to avoid lateral diffusion of the photo-excited carriers. The luminescence spectrum measurement of the LED part shows two peaks located at 873 nm and 889 nm, both of which are caused by recombination from the combined states both in conduction bands and valence bands in the $In_{0.1}Ga_{0.9}As$ quantum well[20].

The optical setup of the up-conversion measurement is shown in Fig.3a. A $CO_2$ laser beam (peak position @ 10.6 μm) was used as the MIR light source. The laser beam is incident on the HIWIP detector on the backside of the up-converter through two KRS5 windows. The NIR light emitted by the LED part transmits through two quartz windows and is collected by a fiber probe/Si photodiode.

The NIR spectra of the LED part induced by up-conversion are shown in Fig.3b&c. For the sake of illustration, the spectra shown here have subtracted the background spectrum caused by the LED itself emitting at different biases. It can be seen that a significant up-conversion process occurs in the device when there is an IR laser irradiation on the surface of the HIWIP-LED sample. The intensity of the two characteristic peaks is positively correlated with both the bias voltage and the incident light power, indicating that the signals are indeed generated by the up-conversion process and that MIR photons are successfully upconverted to NIR photons by the HIWIP-LED device.

Fig.3d indicates the up-conversion efficiency of the HIWIP-LED sample. The efficiency is calculated by dividing the power of the outgoing NIR light by the effective MIR incident light power. Since the $CO_2$ laser spot has a diameter of 4 mm, which is much larger than the device window size (860 μm) and can completely cover the window, we fixed the HIWIP-LED sample at the position where the photocurrent caused by up-conversion is maximum when the laser is turned on. The outgoing light power refers to the net power of the HIWIP-LED sample, which is calculated by subtracting the background light power caused by the LED's own light emission at different bias voltages from the total outgoing light power. The transmissivity of the two quartz windows has been taken into consideration in the up-conversion efficiency calculation.

The up-conversion efficiency is the intrinsic property of the HIWIP-LED sample, therefore, we find that the efficiency hardly changes as the incident power increases, as shown in Fig.3d. An obvious positive correlation is observed between the up-conversion efficiency and the applied biases. Due to the high electrical field induced impact ionization[27] and hot carrier injection effect[28], the responsivity of the HIWIP-LED sample is enhanced under higher reverse biases. The collection efficiency increases with the bias so that the quantum efficiency is also improved under high biases, resulting in high up-conversion efficiency correspondingly. The maximum up-conversion efficiency reaches 0.0034% under 1.7V, which is mainly limited by a low light extraction efficiency (LEE) of the LED part, which is about 2.4%. Designing a proper metasurface on the LED surface may be an effective method to increase LEE and enhance the up-conversion efficiency[29].

The imaging of the laser spot with a wavelength of 10.6 μm has been achieved. The schematic diagram of the optical setup for imaging is shown in Fig.4a. The diameter of the original laser spot is 4 mm, which is too large for the imaging on a single HIWIP-LED window (860 μm×860 μm). Therefore, the laser beam is first focused by a ZnSe lens with a focal length of 63.5 mm and then passes through a KRS5 window with a transmissivity of 74% at 10.6 μm to reach the surface of the HIWIP part of the sample. The image of the laser spot is restored by the LED part and transmitted as an NIR image. Two K9 glass lenses are used to refocus the image onto the CCD so that the up-conversion image can be detected and recorded by the computer.

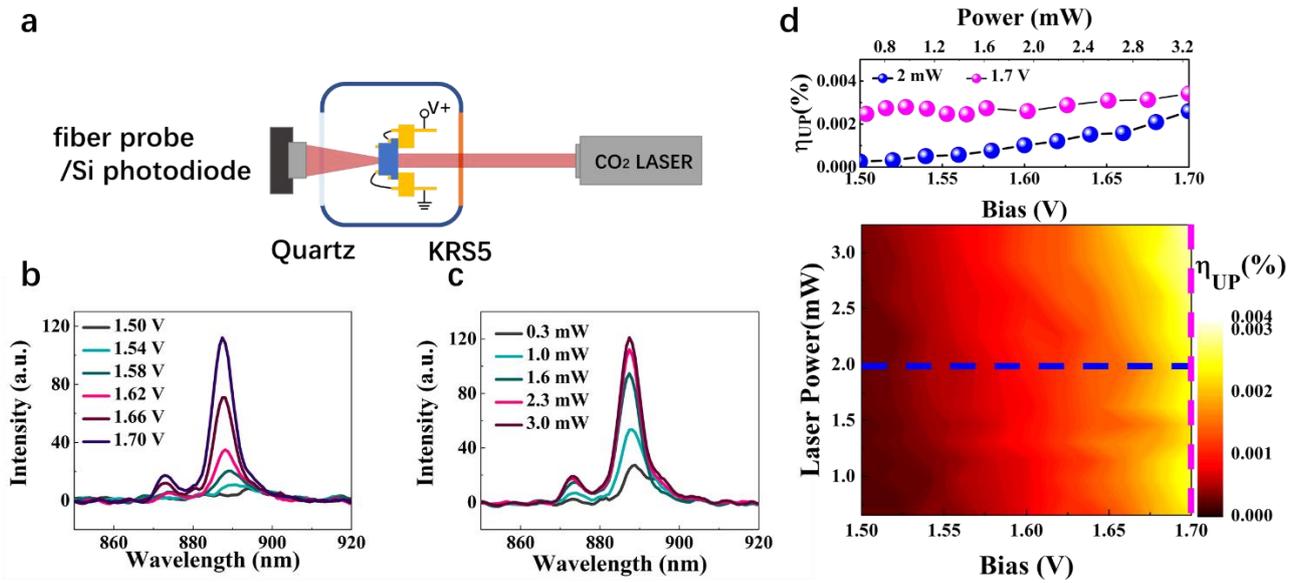

**Fig.3 a** The scheme of the optical setup for up-conversion measurement and the spectrum of the $CO_2$ laser. **b** The 'net' spectrums of the HIWIP-LED under different biases. The incident light power is fixed at 2.0 mW. **c** The 'net' spectrums of the HIWIP-LED under the bias of 1.70 V, with incident light power changing. **d** The up-conversion efficiency of the HIWIP-LED under different incident light power and various biases.

The up-conversion images of the laser spot under different incident light powers and different biases at 8 K are shown in Fig.4b&c. We have taken two sets of images at different biases and different incident light power, respectively. In each set of images, we used the images taken with the $CO_2$ laser off as the background. After subtracting the background from the image taken with the $CO_2$ laser on, a laser spot can be clearly observed, which is recovered by the HIWIP-LED after up-conversion. The intensity of the laser spot on the image is positively correlated with both the incident light power and the bias on the HIWIP-LED sample.

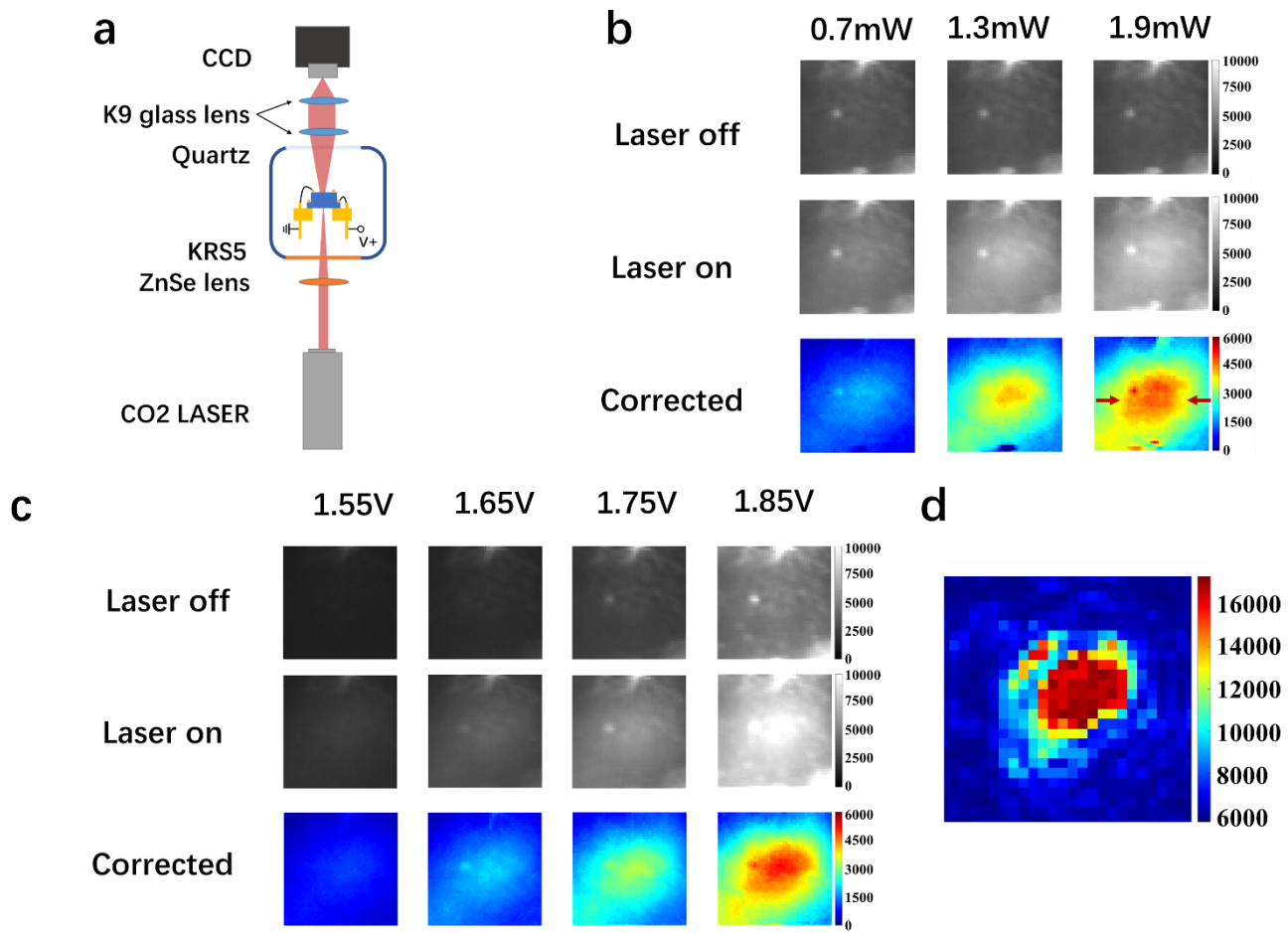

**Fig.4 a** The schematic diagram of the optical setup for MIR up-conversion imaging. **b** The imaging of the laser spot with the incident light power changing. The bias is set as 1.80 V. **c** The imaging of the laser spot under different biases. The incident light power is set as 1.3 mW. **d** The laser spot pictured by a commercial camera.

The photo of the same laser spot is taken by a commercial thermal imaging camera (NEC corporation IRV-T0831C) at the same position (Fig.4d). Both the shapes and the sizes of the laser spots pictured by the up-conversion imaging system and the commercial camera are very similar. Since the HIWIP-LED sample is based on pixelless imaging, in which imaging scheme single cells consisting of a detector and readout circuit is unnecessary, imaging with the HIWIP-LED relies directly on carrier transport in the device so that the image is reproduced in the LED part and output to the Si CCD as

NIR light. In this imaging scheme, the image resolution of the HIWIP-LED sample itself can be very high, and the resolution of the whole imaging system is limited only by the resolution of the Si CCD. Due to the high resolution of the commercial Si CCD, the picture taken by the up-conversion system shows a much higher resolution than the commercial camera. The resolution can also be further improved by optimizing the up-converter or the optical path of the imaging system.

**Response in NIR region**

In addition to the up-conversion function, it is found interesting that the device also shows a response to visible and NIR light at room temperature. We measured the spectra of the HIWIP-LED sample in the NIR region. Fig.5 shows the two spectra that are measured at liquid helium temperature and room temperature, respectively. It is worth mentioning that the HIWIP itself does not show response in the NIR region, which implies that the NIR response is related to the LED part. The upper limit of the NIR spectrum is set to 15000 cm$^{-1}$, which is the maximum value available in the FTIR.

The photoresponse spectra in Fig.5a show that the LED part works as a NIR and visible photovoltaic detector operating up to room temperature. At extreme low temperature, the LED part covers a wavelength region above 11200 cm$^{-1}$, resulting from interband transitions from the GaAs absorption, the In$_{0.1}$Ga$_{0.9}$As quantum well, and the exciton interaction. The photocurrent increases rapidly for photon energies higher than the bandgap energy of the GaAs (1.435 eV) due to absorption[30]. The spike appearing at the onset of the GaAs absorption (peak position at 12220 cm$^{-1}$) is caused by exciton-exciton interaction in GaAs/AlGaAs structure[31]. The long tail from 11940 cm$^{-1}$ to 11210 cm$^{-1}$ is contributed by the confined quantum-well states in both conduction and valence bands ($C_2 \rightarrow HH_2$(838.97 nm), $C_1 \rightarrow LH_1$(876.88 nm), $C_1 \rightarrow HH_1$(885.79 nm)). The schematic diagram of the band structure is shown in the inset of Fig.5a. $C_1$ and $C_2$ are confined states in the conduction bands of the

$In_{0.1}Ga_{0.9}As$ quantum well. Symbols HH and LH stand for heavy hole and light hole. Electrons are excited from the valence band to $C_2$ in the quantum well and easy to escape as $C_2$ is close to the top of the barrier[32]. Transitions to $C_1$ also occur, however, the electrons excited to $C_1$ are more likely to be trapped, and recombine with holes in the quantum well, giving rise to the smaller photocurrent. At room temperature, the response spectra generate a red shift because the bandgap of both the quantum well and the barrier narrows as temperature increases. Since the $In_{0.1}Ga_{0.9}As$ quantum well is very thin, the generated photocurrent is much smaller compared with the GaAs barrier absorption.

The photocurrent decreases substantially after the HIWIP-LED starts to emit light (> 0.8 V) due to the growth of the recombination rate of the carriers in the $In_{0.1}Ga_{0.9}As$ quantum well, as shown in Fig.5b. When the HIWIP-LED is under reverse bias (-0.2V ~ -0.9V), the PN junction in LED is under forward bias and the built-in field is weakened. The energy bands are tilted towards the quantum well so that more photo-generated carriers are injected in. The vast majority of the carriers inside the well stay on the ground states ($C_1$ and $HH_1$), leading to a rapid increase of $C_1$-$HH_1$ transitions. After the reverse bias gets -0.9V, the radiative recombination dominates so that electrons excited from the valence band to the confined states are more likely to recombine with holes rather than contribute to the photocurrent. In contrast, when the HIWIP-LED is under forward bias (0.5V), the LED works as a normal photovoltaic device.

The results demonstrate that the LED part can work as a bi-functional device with the flexibility to switch between emitter and NIR detector by simply changing the bias voltage. More importantly, the response and EL spectra induced by the quantum well show a large overlap, which indicates that the bi-functional LED part would be able to detect the light emitted by an identical device. With the advancement of the semiconductor manufacturing process, similar designs can be applied in various areas like integrated sensing[33] and bidirectional optical communication[34, 35] for the advantages of stability, affordability, and versatility.

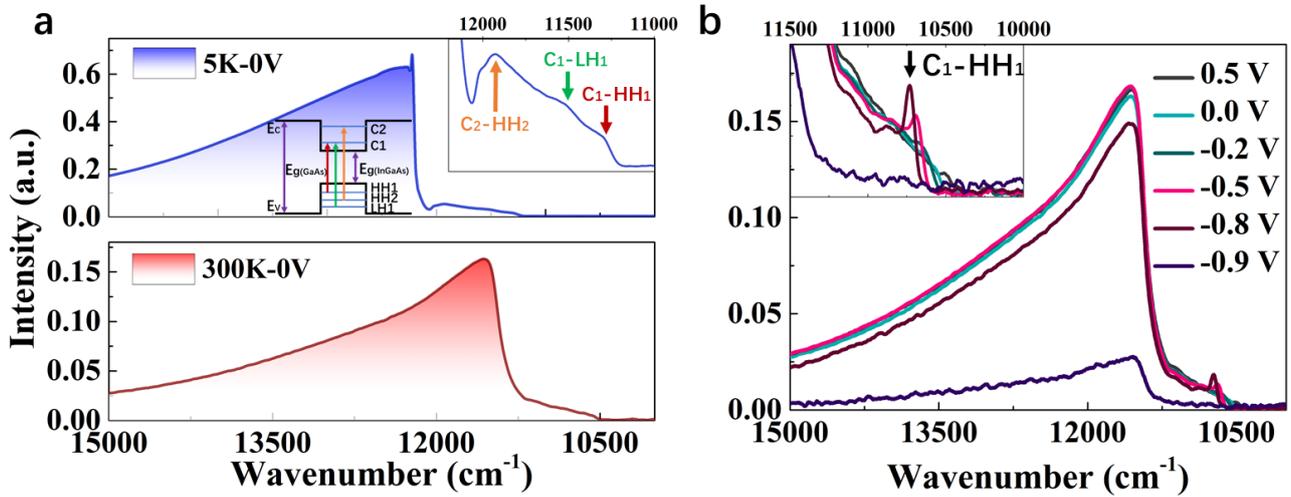

**Fig.5** The NIR spectrum of HIWIP-LED sample in NIR range. **a** The spectrum measured at liquid helium temperature and measured at room temperature under zero bias. **b** The NIR spectrums of the HIWIP-LED sample under different biases at room temperature. The sharp peak caused by the rapid increment of $C_1$-$HH_1$ transition as the reverse bias increases is shown in the inset.

## Discussion

We demonstrated a novel ultra-broadband up-conversion photon-type HIWIP-LED imaging device, which can respond to the radiation from visible to THz region. The responsivity achieves a maximum of 0.14 A/W. The IR up-conversion imaging is realized with a $CO_2$ laser with a peak wavelength of 10.6 μm. The maximum up-conversion efficiency is 0.0034%, which can be further improved by enhancing the light extraction efficiency. The image of the laser spot taken by the up-conversion united system is compared with the image taken by a commercial camera. The up-converter shows higher image quality, which can be further improved by optimizing the up-converter and the optical path. In addition to the up-conversion function, the HIWIP-LED also shows a response to the NIR and visible light at a wavenumber greater than 10200 $cm^{-1}$ at room temperature, indicating the potential bi-functional usage of the proposed up-conversion device.

The HIWIP-LED up-conversion device shows unique advantages to be used in future ultra-broadband imaging systems. First, the united structure of the HIWIP-LED enables a cost-saving fabrication by avoiding the need for extra integrated circuits. Second, the ultra-wide response allows it to be used for versatile analysis, which is useful for systems requiring cost-efficient and compact design like remote sensing and astronomy observations. In addition, the up-converter itself could be looked upon as an ultra-broadband detector covering the whole infrared detection span from visible to THz, indicating that it has great prospects for application in spectrometers if the operating temperature of the device is further increased by designing an appropriate surface structure to enhance optical coupling at a later stage.

The implementation of the ultra-broadband up-conversion devices based on p-GaAs semiconductor HIWIP-LED structure has laid the foundation for the optimal design of photon-type up-conversion devices in the future. Nowadays, GaAs-based photon-type detectors still need to be operated under liquid helium temperature. Finding an effective way to raise the operating temperature is a major direction for future designs. More importantly, the GaAs-based photon-type detectors have great potential for high-speed detection. Therefore, HIWIP-LED is also a strong contender for future high-speed ultra-broadband imaging systems.

## Methods

### Fabrication details

The HIWIP-LED up-converter consists of a p-GaAs HIWIP detector and an AlGaAs/GaAs/In$_{0.1}$Ga$_{0.9}$As quantum well LED directly grown on 600 μm thick semi-insulating GaAs substrates. The HIWIP detector consists of a 20-period homojunction structure, each period containing a 150 Å p-

GaAs emitter layer with a doping concentration of $8\times10^{18}$ cm$^{-3}$ and an 800 Å undoped GaAs barrier layer. The LED part consists of two AlGaAs/GaAs(800 Å/400 Å) heterojunctions and a 90 Å thick In$_{0.1}$Ga$_{0.9}$As quantum well layer sandwiched between them, which is specially designed for operation at liquid helium temperatures. The upper AlGaAs/GaAs heterojunction is highly doped with Si to a concentration of $2.5\times10^{18}$ cm$^{-3}$, forming a p-n junction in the LED part. The HIWIP detector and LED are located between a p-GaAs bottom contact with a doping concentration of $3\times10^{18}$ cm$^{-3}$ and an n-GaAs top contact with a doping concentration of $2.5\times10^{18}$ cm$^{-3}$.

The up-conversion devices were prepared by etching $1\times1$ mm$^2$ mesas using a wet etching technique. Then the Ti\Pt\Au layer (for p-type contact) and Pb\Ge\Ti\Pt\Au layer (for n-type contact) were evaporated onto the bottom contact layer and the top contact layer, respectively, to ensure good ohmic contact. The size of the window reserved for the LED light output on the top contact layer is 860 μm×860 μm. The samples are mounted on 14 pin packages for electrical and optical measurements.

**Measurement details**

The photocurrent spectrums were measured on a Fourier transform infrared spectrometer (Brucker VERTEX 80 IFS 66v/s). The device chosen in the experiment has an active area of 860 μm×860 μm. The spectra on THz/MIR/NIR region were measured with an HDPE window, a KRS5 window, and a quartz window respectively.

The responsivity spectra were measured using a calibrated blackbody (Infrared Systems Development Corporation IR-564/301), a low noise current preamplifier (Model SR570), and a lock-in amplifier (Model SR830).

In the up-conversion experiments, the photoemission spectra of the LED are measured by a fiber spectrometer (Ocean optics QE65PRO). The photoemission power of the LED was measured by

Thorlabs S130C large area Si slim photodiode. The HIWIP-LED was fixed on the sample holder of the cryostat, which was vacuumed to ~$1\times10^{-5}$ mbar and cooled to the liquid helium temperature. The device was irradiated with 10.6 μm light emitted by a $CO_2$ laser. The laser passed through two KRS5 windows and illuminated the active region of the device. The up-conversion device converts the MIR light to NIR light, which passed through the quartz windows on the other side of the cryostat and is detected by a fiber probe or Si photodiode set close to the window.

For the imaging, the diameter of the original laser spot is 4 mm, which is too large for the imaging on a single HIWIP-LED window (860 μm×860 μm). Therefore, the laser beam is first focused by a ZnSe lens with a focal length of 63.5 mm and then passes through a KRS5 window to reach the surface of the HIWIP part of the sample. The image of the laser spot is restored by the LED part and is transmitted as NIR light. Two K9 glass lenses are used to refocus the image onto the CCD (iKon-M 934 BR-DD) so that the up-conversion image can be detected and recorded by the computer.

## Data availability

The data that support the findings of this study are available from the authors on reasonable request, see author contributions for specific data sets.

## References

1. Sizov, F. & Rogalski, A. THz detectors. Prog. Quantum Electron. 34, 278-347 (2010).

2. Tonouchi, M. Cutting-edge terahertz technology. Nat. Photonics 1, 97-105 (2007).

3. Wang, G. Y. et al. Two dimensional materials based photodetectors. Infrared Phys. Technol. 88, 149-173 (2018).

4. Cai, X. et al. Sensitive room-temperature terahertz detection via the photothermoelectric effect in graphene. Nat. Nanotechnol. 9, 814-819 (2014).

5. Chaharmahali, I. & Biabanifard, S. Ultra-broadband terahertz absorber based on graphene ribbons. Optik 172, 1026-1033 (2018).

6. Jabbarzadeh, F. et al. Modification of graphene oxide for applying as mid-infrared photodetector. Appl. Phys. B 120, 637-643 (2015).

7. Zhang, Y. Z. et al. Broadband high photoresponse from pure monolayer graphene photodetector. Nat. Commun. 4, 11 (2013).

8. Viti, L. et al. Efficient Terahertz detection in black-phosphorus nano-transistors with selective and controllable plasma-wave, bolometric and thermoelectric response. Sci. Rep. 6, 10 (2016).

9. Long, M. S. et al. Room temperature high-detectivity mid-infrared photodetectors based on black arsenic phosphorus. Sci. Adv. 3,  (2017).

10. Baeg, K. J. et al. Organic Light Detectors: Photodiodes and Phototransistors. Adv. Mater. 25, 4267-4295 (2013).

11. Burrows, P. E. et al. Reliability and degradation of organic light emitting devices. Appl. Phys. Lett. 65, 2922-2924 (1994).

12. Do, L. M. et al. Observation of degradation processes of Al electrodes in organic electroluminescence devices by electroluminescence microscopy, atomic force microscopy, scanning electron microscopy, and Auger electron spectroscopy. J. Appl. Phys. 76, 5118-5121 (1994).

13. Aziz, H. et al. Degradation processes at the cathode/organic interface in organic light emitting devices with Mg:Ag cathodes. Appl. Phys. Lett. 72, 2642-2644 (1998).

14. Choi, K. K., Allen, S. C., Sun, J. G. & DeCuir, E. A. Resonant detectors and focal plane

arrays for infrared detection. Infrared Phys. Technol. 84, 94-101 (2017).

15. Choi, K. K., Leung, K. M., Tamir, T. & Monroy, C. Light coupling characteristics of corrugated quantum-well infrared photodetectors. IEEE J. Quantum Electron. 40, 130-142 (2004).

16. Zhang, R. et al. Terahertz quantum well photodetectors with reflection-grating couplers. Appl. Phys. Lett. 105, 4 (2014).

17. Luo, H., Liu, H. C., Song, C. Y. & Wasilewski, Z. R. Background-limited terahertz quantum-well photodetector. Appl. Phys. Lett. 86, 3 (2005).

18. Fu, Z. L. et al. Frequency Up-Conversion Photon-Type Terahertz Imager. Sci. Rep. 6, 8 (2016).

19. Dupont, E. et al. Pixelless thermal Imaging with integrated quantum-well infrared photodetector and light-emitting diode. IEEE Photonics Technol. Lett. 14, 182-184 (2002).

20. Bai, P. et al. Broadband THz to NIR up-converter for photon-type THz imaging. Nat. Commun. 10, 3513 (2019).

21. Perera, A. G. U., Yuan, H. X. & Francombe, M. H. Homojunction internal photoemission far-infrared detectors: Photoresponse performance analysis. J. Appl. Phys. 77, 915-924 (1995).

22. Shen, W. Z. et al. Bias effects in high performance GaAs homojunction far-infrared detectors. Appl. Phys. Lett. 71, 2677-2679 (1997).

23. Perera, A. G. U. Heterojunction and superlattice detectors for infrared to ultraviolet. Prog. Quantum Electron. 48, 1-56 (2016).

24. Weerasekara, A. B. et al. Si doped GaAs/AlGaAs terahertz detector and phonon effect on the responsivity. Infrared Phys. Technol. 50, 194-198 (2007).

25. Lao, Y. F. et al. Light-hole and heavy-hole transitions for high-temperature long-wavelength infrared detection. Appl. Phys. Lett. 97, (2010).

26. Bai, P., Zhang, Y. H. & Shen, W. Z. Infrared single photon detector based on optical up-converter at 1550 nm. Sci. Rep. 7, 12 (2017).

27. Shen, W. Z. & Perera, A. G. U. Photoconductive generation mechanism and gain in internal photoemission infrared detectors. J. Appl. Phys. 83, 3923-3925 (1998).

28. Lao, Y. F. et al. Tunable hot-carrier photodetection beyond the bandgap spectral limit. Nat. Photonics 8, 412-418 (2014).

29. Lee, C. W., Choi, H. J. & Jeong, H. Tunable metasurfaces for visible and SWIR applications. Nano Converg. 7, (2020).

30. Panish, M. B. & Casey, H. C. Temperature Dependence of the Energy Gap in GaAs and GaP. J. Appl. Phys. 40, 163-167 (1969).

31. Yaremenko, N. G. et al. Exciton-exciton interaction in GaAs/AlGaAs quantum wells under intense optical excitation. Dokl. Phys. 51, 403-407 (2006).

32. Liu, H. C. et al. GaAs/AlGaAs quantum-well photodetector for visible and middle infrared dual-band detection. Appl. Phys. Lett. 77, 2437-2439 (2000).

33. Jokerst, N. et al. Chip scale integrated microresonator sensing systems. J. Biophotonics 2, 212-226 (2009).

34. Hiraki, T. et al. III-V/Si integration technology for laser diodes and Mach-Zehnder modulators. Jpn. J. Appl. Phys. 58, (2019).

35. Gordon, G. R. J., Howarth, C. & MacVicar, B. A. Bidirectional Control of Blood Flow by Astrocytes: A Role for Tissue Oxygen and Other Metabolic Factors. In: HYPOXIA:

TRANSLATION IN PROGRESS (eds Roach, R. C., Wagner, P. D. & Hackett, P. H.) (2016).

## Acknowledgements

This work was supported by the Natural Science Foundation of China (12074249, 12104061, 11834011, 61974151, 61734006, and 61927813), Natural Science Foundation of Shanghai (19ZR1427000), Project funded by China Postdoctoral Science Foundation (2020M680458), Open Project funded by Key Laboratory of Artificial Structures and Quantum Control (2020-03).

## Author contributions

Y.H.Z conceived the experiment. X.H.L. carried out the whole experiment. X.H.L., P.B., and Y.H.Z. contributed to the data analysis and figures. Z.L.F., D.X.S., Z.Y.T., and J.C.C. contributed to the measurement of the HIWIP detector. S.H.H, X.Q.B., W.J.S., X.R.L., C.H., Z.W.S., C.T., and G.Y.X. contributed to the measurement of the LED and up-converter. X.H.L. wrote the main paper, X.H.L., P.B., Y.H.Z., Z.L.F, and W.Z.S. reviewed the paper. All authors discussed the results and contributed to the paper.

## Competing interests

The authors declare no competing interests.